
\magnification = 1200
\font\titlea=cmb10 scaled\magstep1
\font\titleb=cmb10 scaled\magstep2

\tolerance=10000
\baselineskip=18pt
\rightline{IC/93/179}
\rightline{hepth/9307119}
\vskip .5cm
\centerline{\titleb The Hamiltonian Structures of the Super KP Hierarchy}
\centerline{\titleb Associated with an Even Parity SuperLax Operator}
\vskip 2cm
\baselineskip=14pt
\centerline{ J. Barcelos-Neto\footnote{$^1$}{Parmanent Address: Instituto
de Fisica, Universidade Federal do Rio de Janeiro, Rio de Janeiro, RJ
21945, Brazil.},~~~~Sasanka Ghosh\footnote{$^2$}{On leave from: Institute of
Mathematical Sciences, CIT Campus, Madras 600 113, India.}}
\centerline{and}
\centerline{ Shibaji Roy\footnote{$^3$}{Address after 1st November, 1993:
Institute for Theoretical Physics, Nijenborgh 4, 9747 AG Groningen, The
Netherlands.}}
\bigskip
\centerline{\it International Centre for Theoretical Physics}
\centerline{\it Trieste; Italy.}
\vskip 1cm
\baselineskip=18pt
\centerline{\titlea ABSTRACT}
\bigskip
We consider the even parity superLax operator for the supersymmetric KP
hierarchy of the form $L~=~ D^2 + \sum_{i=0}^\infty u_{i-2} D^{-i+1}$ and
obtain the two Hamiltonian structures following the standard method of
Gelfand and Dikii. We observe that the first Hamiltonian structure is local
and linear whereas the second Hamiltonian structure is non-local and
nonlinear among the superfields appearing in the Lax operator. We discuss
briefly on their connections with the super $w_{\infty}$ algebra.
\vfill

\eject
\baselineskip=18pt
\noindent {\titlea I. Introduction:}
\medskip
One of the most intriguing aspects of the integrable models is the existence
of the biHamiltonian structures [1-3] and their relationship with the extended
conformal algebras [4,5]. The ordinary KP hierarchy [6] formulated in terms
of a Lax
pair is well-known [7,8] to possess a biHamiltonian structure where the first
Hamiltonian structure is isomorphic to the $w_{1+\infty}$ algebra whereas
the second Hamiltonian structure can be identified with the large $N$ limit
of the nonlinear $w_N$ algebra [9-11]. The latter algebra is also known as the
universal ${\hat w}_\infty$ since it contains all possible nonlinear $w_N$
algebras through proper reduction. The existence of the biHamiltonian
structure not only reveals the deep interrelationship between these integrable
models and the underlying conformal field theory but also it enables us to
understand the geometry of the phase-space much better [12]. The free field
realizations of both these abovementioned algebras have been obtained at the
classical as well as at the quantum level in refs.[13,14].

In this letter, we consider a supersymmetric generalization of the ordinary
KP hierarchy in terms of an even parity Lax pair operators. Originally the
supersymmetric KP hierarchy was formulated by Manin and Radul [15]
in terms of an
odd parity Lax operator, but, it is not clear how to take the bosonic limit
of the Hamiltonian structures associated with this Lax operator [16].
As in ref.[17],
we, therefore, consider the even parity Lax operator as the proper
supersymmetric version of the bosonic KP Lax operator and study the Hamiltonian
structures. The first Hamiltonian structure of this Lax operator was first
obtained in ref.[17] using variational technique. We here generalize the
conventional method of Gelfand and Dikii [18,19] to the supersymmetric case and
rederive
the first Hamiltonian structure which precisely matches with the structure
obtained in ref.[17]. The advantage of this approach is that, it automatically
incorporates the requirement that the Hamiltonian structure should satisfy
the Jacobi identity. The bosonic part of this algebra is isomorphic to the
direct sum of two $w_{1+\infty}$ algebra and is therefore the proper
supersymmetric generalization of the bosonic $w_{1+\infty}$ algebra. It also
contains a twisted $N=2$ superconformal algebra [20] as a subalgebra.
We also obtain
the hitherto unknown second Hamiltonian structure associated with this even
parity super Lax operator. Unlike the first Hamiltonian structure it is both
non-local and nonlinear. Under the condition when the bosonic superfields are
set to zero the algebra becomes local. The bosonic part of this algebra when
the fermions are absent becomes identical with the second Hamiltonian
structure of the ordinary KP hierarchy. It also contains another universal
${\hat w}_\infty$ algebra when we look at the algebra of the bosonic part
of the bosonic superfields and other fields are set to zero.

The organization of our paper is as follows. In section II, we fix our
notations
and conventions and explicitly give a few equations of motion following from
the
Lax equation for comparison with the Hamiltonian structures. The derivation of
the first Hamiltonian structure elaborating our method is given in section III.
In section IV, the second Hamiltonian structure is derived and a Lenard type
recursion relation among the two Hamiltonian structures and the two successive
Hamiltonians is given. Finally, we present our concluding remarks in section V.
\vskip 1cm
\noindent {\titlea II. The Lax Operator and the Equations of Motion:}

\medskip
We consider the even parity Lax operator for the super KP hierarchy of the form
$$
L = D^2 + \sum^{\infty}_{i=0} u_{i-2} D^{-i+1} \eqno (2.1)$$
\noindent where the supercovariant derivative is given by~ $D={\partial \over
\partial \theta } + \theta {\partial \over \partial x}$~ satisfying
{}~$D^2={\partial \over \partial x}$~ and $\theta $ is the odd Grassmann
coordinate with $\theta ^2=0$. The formal inverse of $D$ can be written as
{}~$D^{-1}=\theta + {\partial \over \partial \theta }({\partial \over
\partial x})^{-1}$~ such that~ $DD^{-1}=D^{-1}D=1$. The super KP hierarchy can
be written as the Lax equation
$$
{{\partial L} \over {\partial t_n}} = \left[ L^n_+ , L \right] = - \left[
L^n_- , L \right] \qquad {\rm for}~~n=1,2,3,\ldots \eqno (2.2)$$
\noindent Here $t_n$ are the infinite number of evolution
paramaters and $u_i,\quad i=-2,-1,0,1,\ldots $ in $(2.1)$ are the infinite set
of
superfields which are functions of $x,\theta $ and $t_n$'s with the grading
$|u_i|=i+1$ (mod 2). $L_+(L_-)$ refers to the non-negative (respectively
negative) superdifferential part of the formal pseudo-super-differential
operator $L$.

{}From the Lax equation $(2.2)$, it follows that
$$
{{\partial u_{-2}}\over {\partial t_n}} = {{\partial u_{-1}}\over {\partial
t_n}} = 0 \qquad \qquad \forall n \eqno (2.3)
$$
In other words, the superfields $u_{-2}$ and $u_{-1}$ do not have
any dynamics and therefore, we can consistently (without loss of generality)
set them to be zero. In this case, the Lax operator $(2.1)$ becomes constrained
[21] and would have the form
$$
L = D^2 + \sum^{\infty}_{i=0} u_i D^{-i-1} \eqno (2.4)
$$
In order to write down the Lax equation explicitly one needs to use the
Liebnitz rule
$$
D^k f = \sum^{\infty}_{j=0} \left[{k\atop j}\right] (-)^{|f|(k+j)} f^{[j]}
D^{k-j} \eqno (2.5)
$$
where $k$ could be both positive and negative. $|f|$ denotes the grading
of $f$ and $f^{[j]}$ means $(D^j f)$. The superbinomial coefficient
$\left[{k\atop j}\right]$ for $k>0$ is defined as [15]
$$
\left[{k\atop j}\right]=\cases {\left({[{k\over 2}]\atop [{j\over 2}]}\right)
& for $k\ge j$ and $(k,j)\neq (0,1)$ (mod 2)\cr
0 & otherwise\cr}
$$
and for $k<0$
$$
\left[{k\atop j}\right] = (-)^{\left[{j\over 2}\right]} \left[{{-k+j-1}\atop
j}\right] \eqno (2.7)
$$
where $\left[{k\over 2}\right]$ denotes the integral part of ${k\over
2}$ and $\left({m\atop n}\right)$ is the ordinary binomial coefficient.

The first few equations of motion following from $(2.2)$ and $(2.4)$ have the
form
$$
\eqalignno{{{\partial u_i}\over {\partial t_1}}& = u_i^{[2]} & (2.8)\cr
{{\partial u_i}\over {\partial t_2}} & = u_i^{[4]}
+ 2 u_{i+2}^{[2]} + 2
u_0 u_i^{[1]} + 2 ((-)^i - 1) u_0 u_{i+1} \cr
& - 2 \sum^i_{m=0} \left[{{i+1}\atop
{m+1}}\right] (-)^{i+[-{m\over 2}]} u_{i-m} u_0^{[m+1]} - 2 \sum^{i-1}_{m=0}
\left[{i\atop {m+1}}\right] (-)^{[-{m\over 2}]} u_{i-m-1} u_1^{[m+1]}  &
(2.9)\cr
{{\partial u_i}\over {\partial t_3}} & = u_i^{[6]}
+  3 u_{i+2}^{[4]}
 + 3 u_{i+4}^{[2]} + 3 ((-)^i - 1) u_0 u_{i+3} + 3 u_0 u_{i+2}^{[1]}
+ 3 (-)^{i+1} \left[{{i+3}\atop 1}\right] u_0^{[1]} u_{i+2} \cr & + 3 (-)^i
u_0 u_{i+2}^{[2]} + 3 \left[{{i+3}\atop 2}\right] u_0^{[2]} u_{i+1}
+ 3 ((-)^i - 1) u_0^{[2]} u_{i+1} + 3 u_0 u_i^{[3]} + 3 u_0^{[2]} u_i^{[1]}
\cr & + 3 ((-)^i - 1) u_2 u_{i+1} + 3 u_1 u_i^{[2]} + 3 u_1^{[2]} u_i
+ 3 (-)^{i+1} \left[{{i+2}\atop 1}\right] u_1^{[1]} u_{i+1} + 3 u_2 u_i^{[1]}
+ 3 u_3 u_i \cr & + 3 \sum^i_{m=0} (-)^{i+[-{m\over 2}]} \left[{{i+1}\atop
{m+3}}\right] u_{i-m} u_0^{[m+3]} + 3 \sum^i_{m=0} (-)^{[{m\over 2}]}
\left[{i\atop {m+2}}\right] u_{i-m} u_1^{[m+2]} \cr & - 3 \sum^i_{m=0}
(-)^{i+[-{m\over 2}]} \left[{{i+1}\atop {m+1}}\right] u_{i-m} u_2^{[m+1]}
- 3 \sum^i_{m=0} (-)^{[{m\over 2}]} \left[{i\atop m}\right] u_{i-m}
u_3^{[m]} & (2.10)\cr}
$$
In writing down the equation $(2.10)$ we have made use of the identity
$\left[{{m+2}\atop {n+2}}\right] = \left[{{m}\atop {n}}\right] + \left[{m\atop
{n+2}}\right]$. By looking at the equations of motion for the first few
superfields, it is easy to check that unlike in the bosonic KP case [8],
it is not
possible to write down the super KP equation in a local form in terms of one
superfield $u_0$ alone.

By making use of a general theorem for any two pseudo-super-differential
operators $P,Q$ of the form
$$
\int dX~ sRes~ \left[ P , Q \right] = 0 \eqno (2.11)
$$
where the super coordinate $(x,\theta )$ is collectively denoted by
$X$ and `$sRes$' means coefficient of the $D^{-1}$ term (with $D^{-1}$ placed
to
the right), one can show that
$$
H_m = {1\over m} \int dX ~(sRes~ L^m) \eqno (2.12)
$$
$m=1,2,3,\ldots$ are the conserved quantities (Hamiltonians) of the
system which are in involution. The first four Hamiltonians corresponding to
the Lax operator $(2.4)$ are given as
$$
\eqalign{ H_1 & = \int dX u_0(X) \cr H_2 & = \int dX u_2(X) \cr H_3 & =
\int dX \left(u_4(X) + 2 u_0(X) u_1(X) + u_0(X) u_0^{[1]}(X)\right) \cr
H_4 & = \int dX
\left(u_6(X) + 3 u_0(X) u_3(X) + 3 u_1(X) u_2(X) + 3 u_0^{[1]}(X) u_2(X)
\right)} \eqno(2.13)
$$

\vskip 1cm

\noindent{\titlea III. The First Hamiltonian Structure:}

\medskip

The supersymmetric generalization of the first Hamiltonian structure of
Gelfand and Dikii is given by
$$
\{ F_P(L) , F_Q(L) \}_1 = - Tr ([P,Q] L) = -Tr ([L,P] Q) \eqno (3.1)
$$
Here $P$ and $Q$ are some auxiliary operators defined as [19]
$$
P = \sum^{\infty}_{i=-2}D^i p_i , \quad \quad Q = \sum^{\infty}_{i=-2} D^i
q_i \eqno (3.2)
$$
where we take $P,Q$ to have even parity and the grading of the
functions $p_i$, $q_i$ are $|p_i| = |q_i| = i$~(mod 2) and the linear
functional $F_P(L)$ is given as
$$
F_P(L) = Tr (LP) \eqno (3.3)
$$
 ``$Tr$'' is defined as
$$
Tr (LP) = \int dX ~sRes~ (LP) = \int dX \sum^{\infty}_{i=0} (-)^i u_i p_i
\eqno(3.4)
$$
and satisfies the cyclicity property. We notice that because of the
constrained nature of the Lax operator $(2.4)$ where the terms with $D$ and
$D^0$ are absent not all the functions appearing in $P,Q$ are independent. In
fact, two of them are linearly dependent  on others. This can be most
easily understood if we look at the right hand side of $(3.1)$, namely [22],
$$
Tr ([P,L] Q) = Tr ({\partial L\over \partial t} Q) = {\partial \over
\partial t}(F_Q(L)) \eqno (3.5)
$$
Since the commutator $[P,L]$ is coming from the Lax equation, it
cannot contain terms with $D$ and $D^0$. This will give us two constraints
among the functions in $P$. Those two constraints are given by
$$
\sum^{\infty}_{k=0}\left[- p^{[2k+2]}_{2k+1} + \sum^{\infty}_{i=0} \left((-)^i
u_i p^{[2k]}_{2k+i+2} - (-)^{i+k} \left[{{i+k+2}\atop k}\right](u_i
p_{k+i+2})^{[k]}\right)\right] = 0 \eqno (3.6)
$$
and
$$
\sum^{\infty}_{k=0}\left[p^{[k+2]}_k + \sum^{\infty}_{i=0} \left(u_i
p^{[k]}_{k+i+1} + (-)^i \left[{{i+k+1}\atop k}\right] (u_i p_{k+i+1})^{[k]}
\right) \right] = 0 \eqno (3.7)
$$
We note that the left hand side of $(3.1)$ does not contain the functions
$p_{-1}$ and $p_{-2}$ whereas the right hand side does contain them.
So, in order to extract the Poisson brackets among the superfields
consistently one has to eliminate these functions. To demonstrate how it works
we write down $(3.1)$ explicitly,
$$
\eqalign{ & \int dX dY \sum^{\infty}_{i=0} \sum^{\infty}_{j=0}
(-)^{i+j} p_i(X)
\{ u_i(X) , u_j(Y) \}_1 q_j(Y) \cr & = -\int dX \sum^{\infty}_{k=0}p_{-2}
\left[q^{[2k+2]}_{2k+1} - \sum^{\infty}_{i=0}u_i q^{[2k]}_{2k+i+2} +
\sum^{\infty}_{i=0}(-)^{i+k+ik} \left[{{i+k+2}\atop k}\right](u_i
q_{k+i+2})^{[k]}\right] \cr & - \int dX \sum^{\infty}_{k=0} p_{-1}
\left[q^{[k+2]}_k + \sum^{\infty}_{i=0} u_i q^{[k]}_{k+i+1} +
\sum^{\infty}_{i=0} (-)^{i+k+ik} \left[{{i+k+1}\atop k}\right]
(u_i q_{k+i+1})^{[k]}\right] \cr & + \int dX \sum^{\infty}_{i=0}
\sum^{\infty}_{j=0} \left[\sum^i_{k=0} (-)^{i+j+ij+[{k\over 2}]}
\left[{i\atop k}\right] p_i u_{i+j-k} q^{[k]}_j -\sum^j_{k=0}
(-)^{i+j+ik} \left[{j\atop k}\right] p_i (u_{i+j-k} q_j)^{[k]}\right]
} \eqno (3.8)
$$
Since, the terms involving $p_{-1}$ and $p_{-2}$ are multiplied precisely by
the constraints $(3.6)$ and $(3.7)$, they are easily removed. This
demonstrates that the first Hamiltonian structure does not interfere with the
constraints $(3.6)$ and $(3.7)$ in the sense that one would have got the same
structure had one started with the auxiliary operators after removing
$p_{-1}$, $p_{-2}$ in the beginning. We will see below that this is not the
case for the second Hamiltonian structure.

It is now easy to extract the Poisson brackets among the superfields. The
result is given below.
$$
\eqalign{ \{ u_i(X) , u_j(Y) \}_1 = & \left[ \sum^i_{k=0}(-)^{ij+j+[{k\over
2}]} \left[{i\atop k}\right] u_{i+j-k}(X)\right. D^k_X \cr
 & \qquad\qquad\qquad \left. - \sum^j_{k=0}
(-)^{ik+j} \left[{j\atop k}\right] D^k_X u_{i+j-k}(X)
\right] \Delta(X - Y) } \eqno (3.9)
$$
Here $\Delta(X - Y) = \delta(x - y) (\theta - \theta')$. For $H_2,
H_3, H_4$ one can check that the Hamiltonian structure $(3.9)$ correctly
reproduces the equations of motion $(2.8-2.10)$. This structure matches
exactly with the result obtained in ref.[17] using a different method.

The superfields $u_i$ are bosonic for $i =$ odd and are fermionic when $i =$
even. If we expand them as
$$
\eqalign{ u_{2i+1}(x,\theta) & = u^B_{2i+1}(x) + \theta u^F_{2i+1}(x) \cr
u_{2i}(x,\theta) & = u^F_{2i}(x) + \theta u^B_{2i}(x) } \eqno (3.10)
$$
The bosonic fields $u^B_{2i+1}(x)$ and ${\tilde u}^B_{2i}(x) = u^B_{2i}(x) +
u^B_{2i+1}(x)$ satisfy two commuting sets of $w_{1+\infty}$ algebra and so,
the bosonic sector of the first Hamiltonian structure is nothing but
$w_{1+\infty}\oplus w_{1+\infty}$. Also we note that, the algebra among the
superfield components $u^B_1, u^F_1, u^B_2, u^F_2$ is not the $N=2$
superconformal algebra as mentioned in the ref.[17]. However, there is a
twisted $N=2$ superconformal structure where the generators are $T \equiv
-u^B_2$,
$J \equiv -u^B_1$, $G^+ \equiv u^F_1-u^F_2$ and $G^- \equiv u^F_2$.
We, therefore, conclude that
the first Hamiltonian structure of the sKP hierarchy has a twisted $N=2$
superconformal algebra as the subalgebra.

\vskip 1cm

\noindent {\titlea IV. The Second Hamiltonian Structure:}

\medskip

The definition of the second Hamiltonian structure is given as
$$ \{ F_P(L) , F_Q(L) \}_2 = Tr [ \{(LP)_+L - L(PL)_+\}Q ]
 = Tr [ \{L(PL)_- - (LP)_-L\}Q ]  \eqno (4.1)$$
\noindent we will use the second expression since it is more convenient to work
 with. Using a similar argument as given in section III after $(3.4)$ we note
that, the expression $\{L(PL)_- - (LP)_-L\}$ cannot contain terms involving $D$
and $D^0$. As before this requirement gives two constraints involving the
functions in
$P$ of the form,
$$
\eqalign{ \sum^{\infty}_{i=0} & \left[ (-)^k u_k p_k - \sum^k_{m=0}(-)^{k(m+1)}
\left[{k\atop m}\right] (p_k u_{k-m})^{[m]} \right] - p^{[2]}_{-1} = 0 \cr
\sum^{\infty}_{i=0} & \left[ u_{k+1} p_k
+ u_k p^{[1]}_k - \sum^k_{m=0}(-)^{m(k+1)} \left[{k\atop m}\right] (p_k
u_{k-m+1})^{[m]} \right] + 2 u_0
p_{-1} + p^{[2]}_{-2} + p^{[3]}_{-1} = 0 } \eqno (4.2)$$
\noindent One can write down the above two contraints in a more compact form
as follows,
$$
\eqalign{ p^{[2]}_{-1} & = sRes~[ L , {\hat P}_0 ] \cr p^{[2]}_{-2} & = -
sRes~([ L , {\hat P}_{-1} ] D) } \eqno (4.3)$$
\noindent where ${\hat P}_{-m} = \sum^{\infty}_{i=-m} D^ip_i$.

Again we note that the left hand side of $(4.1)$ does not contain the
functions $p_{-2}$ and $p_{-1}$, but the right hand side does.
Writing down $(4.1)$ explicitly we find
$$
\eqalign{ & \int dX dY \sum^{\infty}_{i=0} \sum^{\infty}_{j=0} (-)^{i+j}
p_i(X) \{ u_i(X) , u_j(Y) \}_2 q_j(Y) \cr
& = \int dX~ p_{-2}~ sRes~[ L , {\hat Q}_0 ] - \int dX~ p_{-1}~ sRes~
([ L , {\hat Q}_0 ] D) \cr
& + \int dX \sum^{\infty}_{i=0}
\sum^{\infty}_{j=0} \left[ \sum^{j+2}_{m=0} (-)^{m(i+j+1)+i+j+1}
\left[{{j+2}\atop
 m}\right] p_i (u_{i+j-m+2} q_j)^{[m]}\right. \cr
& + \sum^{i+2}_{m=0}
(-)^{j(i+1)+im+i+[-{m\over 2}]} \left[{{i+2}\atop m}\right] p_i u_{i+j-m+2}
q_j^{[m]}  \cr & +
\sum^{j-1}_{k=0} \sum^{j-k-1}_{m=0} (-)^{(i+k)m+ik+jm+j+1} \left[{{j-k-1}\atop
m}\right] p_i u_k (u_{i+j-k-m-1} q_j)^{[m]} \cr
& \left. + \sum^i_{m=0}
\sum^{j-1}_{k=0} \sum^{j-k-1}_{l=0} (-)^{j(i+1)+i(m+k+l)+kl+[-{m\over 2}]}
\left[{i\atop m}\right] \left[{{k+l}\atop l}\right] p_i u_{i+j-m-k-l-1} (u_k
q_j)^{[m+l]}\right] } \eqno (4.4)
$$
In order to derive $(4.4)$ we have made use of the following identites of the
superbinomial coefficients,
$$
\left[{{m+i+2}\atop m}\right] = \sum^m_{l=0} \left[{{l+i}\atop l}\right]
\eqno (4.5)
$$
with $m + l =$ even number,
$$
\left[{{m+i+1}\atop m}\right] = \sum^m_{l=0} (-)^{(m+l)(i+1)}
\left[{{i+l}\atop l}\right] \eqno (4.6)$$
and
$$
\left[{{m+1}\atop {n+1}}\right] = \left[{{m}\atop {n}}\right] + (-)^{n+1}
\left[{m\atop {n+1}}\right] \eqno (4.7)
$$
We note that, in this case, the terms involving $p_{-1}$ and $p_{-2}$
in $(4.4)$
do not vanish as happened in deriving the first Hamiltonian structure. We have
to eliminate these functions in order to extract the Poisson brackets of the
superfields consistently. This can be done using the constraints $(4.2)$.
This, in fact, is the origin of the non locality of the second Hamiltonian
structure. We here give the result,
$$
\eqalign { & \{ u_i(X) , u_j(Y) \}_2 \cr & = \left[ \sum^{j+2}_{m=0}
(-)^{m(i+j)+j+m+1} \left[{{j+2}\atop m}\right] D^m_X u_{i+j-m+2}(X)\right.
\cr &
+ \sum^{i+2}_{m=0} (-)^{j(i+1)+im+[-{m\over 2}]} \left[{{i+2}\atop m}\right]
u_{i+j-m+2}(X) D^m_X \cr
& + \sum^{j-1}_{k=0} \sum^{j-k-1}_{m=0}
(-)^{(i+k)m+ik+jm+i+j+1} \left[{{j-k-1}\atop m}\right] u_k(X) D^m_X
u_{i+j-k-m-1}(X)\cr
&+\sum^i_{m=0}\sum^{j-1}_{k=0}\sum^{j-k-1}_{l=0}
(-)^{j(i+1)+i(m+k+l+1)+kl+[-{m\over 2}]} \left[{i\atop m}\right]\cr
& \qquad\qquad\qquad\qquad
\left[{{k+l}\atop l}\right] u_{i+j-m-k-l-1}(X) D^{m+l}_X u_k(X) \cr
&
+ \sum^{i-1}_{l=0} \sum^j_{m=1} \left[{i\atop {l+1}}\right]
\left[{j\atop m}\right] (-)^{il+j(m+1)+m+[{l\over 2}]+1} u_{i-l}(X)
D_X^{m+l-1} u_{j-m}(X) \cr & + \sum^i_{l=1} \sum^{j-1}_{m=0}
(-)^{il+jm+[-{l\over 2}]} \left[{i\atop l}\right] \left[{j\atop {m+1}}\right]
u_{i-l}(X) D_X^{m+l-1} u_{j-m}(X) \cr & + \sum^i_{l=1} \sum^j_{m=1}
(-)^{il+j(m+1)+m+[-{l\over 2}]} \left[{i\atop l}\right] \left[{j\atop m}\right]
u_{i-l}(X) D_X^{m+l-1} u_{j-m}(X) \cr & + \sum^j_{m=1} (-)^{j(m+1)+m+1}
(1 + (-)^{i+1}) \left[{j\atop m}\right] u_{i+1}(X) D_X^{m-2} u_{j-m}(X) \cr &
+ \sum^i_{m=1} (-)^{im+[-{m\over 2}]+1} (1 + (-)^{j+1}) \left[{i\atop m}\right]
u_{i-m}(X) D_X^{m-2} u_{j+1}(X) \cr & + \sum^i_{m=1} (-)^{im+[-{m\over 2}]+1}
\left[{i\atop m}\right] u_{i-m}(X) D_X^{m-2} u_j(X) D_X \cr & + \sum^j_{m=1}
(-)^{j(m+1)+m+i+1} \left[{j\atop m}\right] D_X u_i(X) D_X^{m-2} u_{j-m}(X) \cr
 & \left. + 2 \sum^i_{l=1} \sum^j_{m=1} (-)^{il+j(m+1)+m+[-{l\over 2}]}
\left[{i\atop l}\right] \left[{j\atop m}\right] u_{i-l}(X) D_X^{l-2} u_0(X)
D_X^{m-2} u_{j-m}(X) \right] \Delta(X-Y) } \eqno (4.8)
$$
One can check that $(4.8)$ precisely reproduces the equations of motions
$(2.8 - 2.10)$ with the Hamiltonians $H_1$, $H_2$ and $H_3$ given in $(2.13)$.

A few remarks are in order here. First of all, we note that the algebra is
nonlinear. It contains terms which are linear, bilinear and trilinear in the
superfields. Secondly, from eighth term onwards, the expression is non-local.
The non-locality, however, disappears when both $i,j$ are even. In this case,
$u_i, u_j$ are fermionic superfields. In the limit when fermionic components
of these fermionic superfields are set to zero, algebra precisely matches with
the second Hamiltonian structure of the ordinary KP hierarchy.
The algebra also becomes
local if we set $u_{2i} = 0$ {\it i.e.} when $i,j$ both are odd. In this case,
the bosonic component of the bosonic superfields also satisfy the same algebra
 as the ordinary KP hierarchy. So, only in these two limits the second
Hamiltonian structure becomes universal ${\hat w}_{\infty}$ algebra.

Since both the Hamiltonian structures $(3.9)$ and $(4.8)$ reproduce the same
Lax equation with two successive Hamiltonians, they would satisfy a Lenard
type recursion relation of the form
$$\int dY \sum_j \{ u_i(X) , u_j(Y) \}_1 {{\delta H_{n+1}}\over
{\delta u_j(Y)}} = \int dY \sum_j \{ u_i(X) , u_j(Y) \}_2 {{\delta H_n}\over
{\delta u_j(Y)}} \eqno (4.9)$$
with $n=1,2,\ldots$. One can alternatively construct the infinite
number of conserved quantites recursively using relation $(4.9)$. This type of
relation is also useful in order to understand the geometry of the phase
space. We also like to remark that the non-locality of the Hamiltonian
structure can be absorbed in the Hamiltonian itself. In that case,
Hamiltonians would become non-local and cannot be obtained from the Lax
operator as in $(2.12)$, but the Hamiltonian structure in this way can be made
local.

\vskip 1cm

\noindent {\titlea V. Concluding Remarks:}

\medskip

The even parity superLax operator may be constructed from the Manin-Radul odd
 parity superLax operator in three different ways
$$\eqalignno{ L & = L^2_{MR} & (5.1a)\cr
 L & = DL_{MR} & (5.1b)\cr
 L & = L_{MR}D & (5.1c)\cr}
$$
where $L_{MR} = D + \sum^{\infty}_{i=0} u_{i-2} D^{-i}$. The
operator $(5.1a)$ mixes the superfields in a complicated way and the
Hamiltonian
structure associated with it is not very illuminating. The super Lax operator
we have considered here is of the type $(5.1c)$. In component form it can be
written as
$$
\eqalign{L = & \left( \partial + \sum^{\infty}_{i=0} u^B_{2i+1}
\partial^{-i-1}\right)
 + \theta
\sum^{\infty}_{i=0} u^F_{2i+1} \partial^{-i-1} - \theta \sum^{\infty}_{i=0}
u^F_{2i} \partial^{-i} \cr & \qquad \qquad
+ \sum^{\infty}_{i=0} u^F_{2i} \partial^{-i-1}
\partial_{\theta} + \sum^{\infty}_{i=0} u^B_{2i} \partial^{-i-1} \theta
\partial_{\theta}} \eqno (5.2)
$$
In the bosonic limit of this operator only $u^B_{2i+1}$ terms
survive, but
we note that in the first Hamiltonian structure we had identified $u^B_2$ as
the energy momentum tensor which are not present in the bosonic limit.
Although $u^B_{2i+1}$ also satisfy $w_{1+\infty}$ algebra but it does not
contain the right energy momentum tensor which gave rise to the twisted $N=2$
superconformal algebra. However, the operator $DL_{MR}$ in the component form
contains both types of bosons in the bosonic limit. Since,
$$
\eqalign{ & DL_{MR} \cr = & \left( \partial + \sum^{\infty}_{i=0} (u^B_{2i} +
u^B_{2i+1})
\partial^{-i-1} \right) + \theta \sum^{\infty}_{i=0} ({u^F_{2i}}'
+ u^F_{2i+1})
\partial^{-i-1} - \theta \sum^{\infty}_{i=0} (u^F_{2i-1} - u^F_{2i})
\partial^{-i} \cr
& \qquad + \sum^{\infty}_{i=0} (u^F_{2i-1} - u^F_{2i})
\partial^{-i-1}
\partial_{\theta} + \sum^{\infty}_{i=0} ({u^B_{2i-1}}' - u^B_{2i})
\partial^{-i-1} \theta \partial_{\theta } } \eqno (5.3)
$$
where, `$'$' denotes the derivative with respect to $x$, we have the terms
involving $(u^B_{2i} +
u^B_{2i+1})$ in the bosonic limit. If $u^B_{2i+1}$ are constructed from the
fermion bilinears and
$u^B_{2i}$ are constructed out of both boson and fermion bilinears as in
ref.[23],
then only bosons that will survive are $u^B_{2i}$ in the limit. In this case,
we will have both $w_{1+\infty}$ algebra and a twisted $N=2$ superconformal
algebra satisfied by $J \equiv - (u^B_0 + u^B_1)$, $G^+ \equiv ({u^F_0}'
+ u^F_2)$,
$G^- \equiv (u^F_1 - u^F_2)$ and $T \equiv (u^B_2 - {u^B_1}')$.
The details of this work will be reported elsewhere.

To conclude, we have derived the first and the second Hamiltonian structures
associated with even parity superLax operator of the sKP hierarchy. The
Poisson brackets among the superfields are both antisymmetric and satisfies
the Jacobi identity as they are obtained from the conventional method of
Gelfand and Dikii in the supersymmetric case. The bosonic part of the first
Hamiltonian structure is isomorphic to $w_{1+\infty}\oplus w_{1+\infty}$ and
contains a twisted $N=2$ superconformal algebra as a subalgebra. The second
Hamiltonian structure is nonlinear and non-local. When the fermionic components
 of the superfields are set to zero, they give rise to two sets of universal
${\hat w}_{\infty}$ algebra.
\vskip 1cm
\noindent {\titlea Acknowledgements:}
\medskip
The authors would like to thank Professor A. Salam, the International Atomic
Energy Agency and UNESCO for hospitality and support at the International
Centre for Theoretical Physics, Trieste.
\vskip 1cm

\noindent {\titlea References:}

\medskip

\item{1.} A. C. Newell, Solitons in Mathematical Physics, SIAM Philadelphia
(1985).

\item{2.} L. D. Faddeev and L. A. Takhtajan, Hamiltonian Methods in the
Theory of Solitons, Springer-Verlag, New York/Berlin (1987).

\item{3.} A. Das, Integrable Models, World Scientific, Singapore (1989).

\item{4.} J. L. Gervais and A. Neveu, Nucl. Phys. B209 (1982) 125.

\item{5.} J. L. Gervais, Phys. Lett. B160 (1985) 277,279.

\item{6.} E. Date, M. Kashiwara, M. Jimbo and T. Miwa in ``Nonlinear Integrable
Systems" eds. M. Jimbo and T. Miwa, World Scientific, Singapore (1983).

\item{7.} F. Yu and Y. S. Wu, Nucl. Phys. B373 (1992) 713.

\item{8.} A. Das, W.-J. Huang and S. Panda, Phys. Lett. B271 (1991) 109.

\item{9.} K. Yamagishi, Phys. Lett. B259 (1991) 436.

\item{10.} F. Yu and Y. S. Wu, Phys. Lett. B263 (1991) 436.

\item{11.} L. A. Dickey, Ann. New York Academy of Sciences (1987) 131.

\item{12.} S. Okubo and A. Das, Phys. Lett. B209 (1988) 311; A. Das and
S. Okubo, Ann. Phys. 190 (1989) 215.

\item{13.} E. Bergshoeff, C. N. Pope, L. J. Romans, E. Sezgin and X. Shen,
Phys. Lett. B245 (1990) 447; E. Bergshoeff, B. deWit and M. Vasiliev, Phys.
Lett. B256 (1991) 199.

\item{14.} I. Bakas and E. Kiritsis, Int. Jour. Mod. Phys. A7 (1992) 55.

\item{15.} Yu. I. Manin and A. O. Radul, Comm. Math. Phys. 98 (1985) 65.

\item{16.} S. Panda and S. Roy, Phys. Lett. B291 (1992) 77.

\item{17.} F. Yu, Nucl. Phys. B 375 (1992) 173.

\item{18.} I. M. Gelfand and L. A. Dikii, Russ. Math. Surv. 30 (1975) 77;
Funct. Anal. and Appl. 10 (1976) 4; ibid 11 (1977) 93.

\item{19.} V. G. Drinfeld and V. V. Sokolov, Jour. Sov. Math. 30 (1985) 1975.

\item{20.} T. Eguchi and S.-K. Yang, Mod. Phys. Lett. A5 (1990) 1693.

\item{21.} In a simpler situation such a constrained Lax operator has been
treated in J. Barcelos-Neto and A. Das, Jour. Math. Phys. 33 (1992) 2743.

\item{22.} For clarification on this point see A. Das and W.-J. Huang,
Jour. Math. Phys. 33 (1992) 2487.

\item{23.} The 1st reference in [13]
\vfill\eject\end